\documentclass[fleqn,usenatbib]{mnras}
\usepackage{amsmath}
\usepackage{xcolor, graphicx}
\usepackage{amssymb}
\usepackage{soul}
\usepackage{txfonts}
\usepackage{caption}

\begin{document}

\title[GC system size]{How large are the globular cluster systems of early-type galaxies and do they scale with galaxy halo properties?}
\author[D. A. Forbes]{Duncan A. Forbes$^{1}$\thanks{E-mail:
dforbes@swin.edu.au}, 
\\
$^{1}$Centre for Astrophysics \& Supercomputing, Swinburne
University, Hawthorn VIC 3122, Australia\\
}


\pagerange{\pageref{firstpage}--\pageref{lastpage}} \pubyear{2002}

\maketitle

\label{firstpage}

\begin{abstract}

The globular cluster systems of galaxies are well-known to extend to large galactocentric radii. Here we quantify the size of GC systems using the half number radius of 22 GC systems around early-type galaxies from the literature. 
We compare GC system sizes to the sizes and masses of their host galaxies. We find that GC systems typically extend to 4$\times$ that of the host galaxy size, however this factor varies with galaxy stellar mass from about 3$\times$ for M$^{\ast}$ galaxies to 5$\times$ for the most massive galaxies in the universe. The size of a GC system scales approximately linearly with the virial radius (R$_{200}$) and with the halo mass (M$_{200}$) to the 1/3 power. The GC system of the Milky Way follows the same relations as for early-type galaxies. For Ultra Diffuse Galaxies their GC system size scales with halo mass and virial radius as for more massive, larger galaxies. UDGs indicate that the linear scaling of GC system size with stellar mass for massive galaxies flattens out for low stellar mass galaxies.  Our scalings are different to those reported recently by Hudson \& Robison (2017). 

\end{abstract}


\section{Introduction}

Globular clusters (GC) can be traced to relatively large galactocentric radii, thus providing valuable probes of their host galaxy halos  
(where the underlying starlight has a low surface brightness). 
This property has been
exploited by the SLUGGS survey of 25 nearby early-type galaxies
(Brodie et al. 2014 and see sluggs.swin.edu.au) 
 to investigate the structural properties (Kartha et al. 2016), 
metallicity (Usher et al. 2012), kinematics (Pota et al. 2013) and dynamical mass (Alabi et al. 2017) of GC systems over a range of host galaxy properties.  

A number of scaling relations have been found between GC systems and their host galaxy. Perhaps the most remarkable is the scaling between the total mass of a GC system and the  host galaxy's halo mass (Blakeslee et al. 1997; Spitler \& Forbes 2009; Georgiev et al. 2010; Hudson et al. 2014; 
Harris et al. 2015; Harris et al. 2017a). This near linear relation holds over a large range in halo mass with little, or no, dependence on host galaxy type.


How far do globular cluster systems extend relative to their host galaxy and do they scale with host galaxy halo properties? 
Globular clusters have been confirmed out to more than 
30 times the effective (half-light) radius of their host galaxy (e.g. Alabi et al. 2016).
However defining the total radial extent of a GC system is problematic.  The total radial extent of a GC system is
usually defined to be the radius at which the number density of GCs
per unit area, from a photometric study, decreases to a constant
level, indicating that a `background'  has been reached. This
constant density background is assumed to be due to contaminants in
the photometric object list.
This approach has been taken by Rhode et al. (2007, 2010) and Kartha et al. (2016). 

However, the level of the background used is somewhat dependent on the ability of the
photometry to separate bona fide GCs from contaminants. For example,
imaging from the ACS camera onboard HST can sufficiently resolve individual GCs
to measure their size 
to distances 
of about 20 Mpc thus reducing contaminant levels to a bare minimum. At the
other extreme, ground-based imaging under poor seeing conditions will
result in object lists that may include significant contributions from
foreground stars and distant galaxies.
A better approach is to measure the effective radius of both the GC system and its host 
galaxy. Such measures are derived from intermediate radial scales, which are relatively less affected by contamination. 

Recently Hudson \& Robison (2017; hereafter HR17) investigated the size of GC systems and trends with halo properties. 
They selected GC candidates around 9 galaxies based on their size, i band magnitude and g--i colour from the wide-field CFHT Lens Survey (Hudelot et al. 2012). 
They fit Sersic profiles to the GC radial surface density profile (fixing the Sersic n parameter to be 4, i.e. a de Vaucouleurs profile) to derive the half number radius (hereafter GC R$_e$) of the GC system. HR17 noted that this approach gave good fits to the GC density profiles but on average their GC R$_e$ sizes were systematically larger than those in the literature (as they fit only in the outer regions of GC systems). HR17 also measured GC R$_e$ for 26 other GC systems using data from the literature, including some measurements of the GC R$_e$ from the SLUGGS survey (Kartha et al. 2014; 2016).


Here we investigate the central question of how large are GC systems and how they scale with host galaxy properties.  
We take measured sizes of GC systems around early-type galaxies (ETGs) using available data from the literature. 
The GC system imaging for these studies comes from 
HST and/or wide-field deep ground-based observations. 
We include the ETGs listed in HR17 and four very massive ETGs galaxies from Harris (2017b) which extends the analysis to the highest mass galaxies in the universe. 
Although not ETGs, it is interesting to examine the GC systems of the new class of galaxy dubbed Ultra Diffuse Galaxies (UDGs). Such galaxies have stellar masses similar to dwarf galaxies of around 10$^8$ M$_{\odot}$ but halo masses closer to giant galaxies (van Dokkum et al. 2017). 

\begin{table}
\centering
{\small \caption{Galaxy and globular cluster system properties}}
\begin{tabular}{@{}ccccccc}
\hline
\hline
Galaxy & Type & Dist. & log M$_{\ast}$ & R$_e$ & GC R$_e$ & Ref\\
 & & [Mpc] &  [M$_{\odot}$] & [kpc] & [kpc]  &\\
(1) & (2) & (3) & (4) & (5)  & (6)\\
\hline
 N720 & E5 & 26.9 & 11.27 & 3.8 & 13.7 (2) & 1\\
N1023 & S0 & 11.1 & 10.99 & 2.6  & 3.3 (0.9) & 1\\
N1407 & E0 &  26.8 & 11.60 & 12.1 & 25.5 (1) & 2\\
N2768 & E6/S0 & 21.8 & 11.21 & 6.4 & 10.6 (2) &1\\
N3607 & S0 & 22.2 & 11.39 & 5.2 & 14.2 (2) & 3\\
N3608 & E1-2 & 22.3 & 11.03 & 4.6 & 9.1 (1) & 3\\
N4278 & E1-2 & 15.6 & 10.95 & 2.1 & 11.3 (2) & 4\\
N4365 & E3 & 23.1 & 11.51 & 8.7 & 41.3 (8) & 5\\
\hline
N4406 & E3 &  17.9 & 11.47 & 8.1& 28.2 (1) & 6\\
N4472 & E2 & 16.7 & 11.83 & 7.7 & 58.4 (8) & 6\\
N4594 & Sa &  9.5 & 11.41 &  3.3 & 16.8 (1) & 6\\
N5813 & E1-2 & 31.3 & 11.43 & 8.7 & 36.6 (3) & 6\\
N4874 & cD & 100 & 11.90 & 15.8 & 62 (2) & 7\\
N4889 & cD & 100 & 12.09 & 16.3 & 110 (--) & 8\\
U9799 & E & 150 & 12.00 & 22.7  & 61 (--) & 8\\
U10143 & cD & 154 & 11.73 & 23.4 & 114 (--) & 8\\
\hline
N3384 & S0 & 10.9 & 10.61 & 1.7 & 7.3 (4.8) & 9\\
N4486 & cD & 16.0 & 11.74 & 6.9 & 87 (56) & 9\\
N4754 & S0 & 16.1 & 10.68 & 2.5 & 8.8 (3.5) & 9\\
N4762 & S0 & 15.3 & 10.67 & 3.3 & 4.7 (1.1) & 9\\
N5866 & S0 & 11.7 & 10.83 & 1.7 & 8.5 (1.8) & 9\\
N7332 & S0pec & 13.2 & 10.25 & 1.9 & 1.4 (0.3) & 9\\
\hline
IC219 & E & 72.7 & 10.97 & 1.88 & 25.9 (8.8) & 9\\
N883 & S0 & 72.7 & 11.40 & 3.83 & 178 (72) & 9\\
N2695 & S0 & 26.5 & 10.54 & 1.15 & 10.3 (6.1) & 9\\
N2698 & S0 & 26.5 & 10.54 & 0.88 & 10.4 (9.9) & 9\\
N2699 & E & 26.5 & 10.30 & 0.79 & 2.0 (1.6) & 9\\
N5473 & S0 & 26.2 & 10.72 & 1.58 & 1.56 (2.4) & 9\\
N5485 & S0 & 26.2 & 10.70 & 2.00 & 12.9 (7.3) & 9\\
\hline
DF17 & UDG & 100 & 7.92 & 3.4 & 5.8 (1.0) & 10\\
DF44 & UDG & 100 & 8.43 & 4.7 & 10.3 (--) & 11\\
DFX1 & UDG & 100 & 8.26 & 3.5 & 7.7 (--) & 11\\
\hline
\end{tabular}
\begin{flushleft}
{\small 

Notes: columns are (1) galaxy name where N=NGC and U=UGC,  
(2) Hubble type, (3)  distance,  (4) stellar mass, (5) galaxy effective radius,  (6) globular cluster system effective radius and uncertainty, (7) globular cluster system reference 1= Kartha et al. (2014), 2=Spitler et al. (2012), 3=Kartha et al. (2016), 4=Usher et al. (2013), 5=Blom et al. (2012), 6=Hargis \& Rhode (2014), 7=Peng et al. (2011), 8=Harris et al. (2017b), 9 =  
Hudson \& Robison (2017; HR17), 10 = Peng et al. (2016), 11 = van Dokkum et al. (2017). 
The table is divided into five sections: SLUGGS survey galaxies, galaxies from Hargis \& Rhode (2014), galaxies from Harris et al. (2017), literature galaxies with GC systems fit by HR17, new data from HR17 and Ultra Diffuse Galaxies in the Coma cluster. Stellar masses and galaxy 
effective are taken from Forbes et al. (2017),  Cappellari et al. (2011), Harris et al. (2017), HR17, 
Veale et al. (2017), Vika et al. (2012), Peng et al. (2016) and van Dokkum et al. (2017). When the GC effective radii R$_e$ 
uncertainty is not quoted, we  assume 10\%. 
}

\end{flushleft}
\end{table}

\section{The sample}

Our sample consists of GC systems of ETGs  using inhomogeneous data from the available literature. 
We exclude the GC systems of late-type galaxies as they tend to be GC-poor, lacking well-defined system sizes. Our ETG sample is 
8 from the SLUGGS survey, 4 from Hargis \& Rhode (2014) including the massive bulge galaxy NGC 4594 (the Sombrero), 6  GC systems fit by HR17 (but based on data from Young et al. 2012 and Hargis \& Rhode 2012), and 7 new ETGs from HR17
(we exclude the interacting pair NGC 942+943 but do include the elliptical galaxy NGC 2699 which was incorrectly identified as an Sb in HR17). All of these galaxies are listed in HR17. To this sample we include four massive, central dominant ETGs studied by Harris et al. (2017b). 
Relevant properties of the sample galaxies and references are given in Table 1. 
We also list in Table 1 three  Ultra Diffuse Galaxies located in the Coma cluster, whose GC systems have recently been studied by Peng et al. (2016) and van Dokkum et al. (2017). 
For comparison purposes we show the Milky Way's GC system in the figures that follow. The Milky Way GC system has a GC R$_e$ of 4.1 kpc (as measured by HR17).  The Milky Way itself has a total stellar mass of log M$_{\ast}$ = 10.81 (Mcmillan 2011) and an effective radius of  2.7 kpc (Gilmore et al. 1989). 


\section{Results and Discussion}

\begin{figure}
      \includegraphics[angle=-90,width=0.5\textwidth]{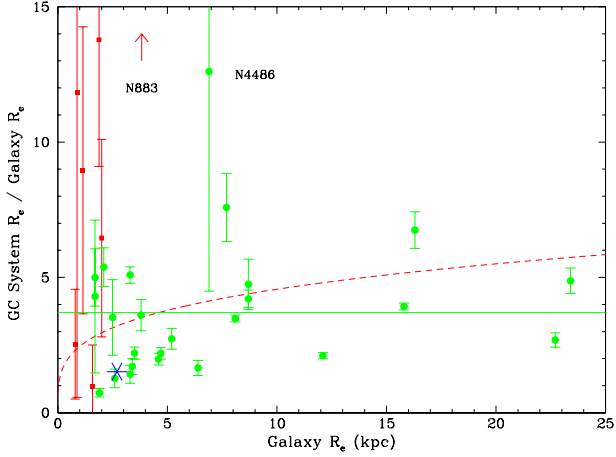}
	    \caption{\label{fig:corrGC2} The ratio of the size of a globular cluster system to its host galaxy vs galaxy size.  
The red filled squares are galaxies from the study of HR17; 
NGC 833 which a ratio of 46 is shown as a lower limit. The green filled circles are other data from the existing literature, including 3 Ultra Diffuse Galaxies. The blue asterisk represents the Milky
Way's GC system.  The UDGs and the Milky Way follow the general trend. Excluding the HR17 data and NGC 4486 (M87), the mean ratio for early-type galaxies  is 3.7 (shown by a green solid line) with a mild trend for an 
increasing ratio in larger galaxies. The relation found by HR17 for their sample of 35 early and late-type galaxies is shown by the red dashed line.}

\end{figure} 

Before investigating the scaling of GC system size with halo properties, we examine host galaxy stellar properties, i.e. size and mass.
In Figure 1 we show the relative size of a GC system (GC R$_e$) to its host
galaxy size (R$_e$).  In 
general, the two measurements for an individual galaxy are carried out
by different studies. The uncertainty in the size ratio shown is only that for the GC R$_e$ as this tends to dominate over the 
quoted uncertainties in the galaxy R$_e$.
The 7 new galaxies from HR17 are highlighted in Fig. 1. They tend to have much larger quoted uncertainties than the existing literature data, and in the case of NGC 883 it lies off the plot due to its large ratio (i.e. GC R$_e$ / R$_e$ = 46) which we suspect to be a combination of an underestimated galaxy R$_e$ (e.g. its location in the galaxy size-mass plot shown in figure 5 of Forbes et al. 2017) and an overestimated GC R$_e$ (see Fig. 2). 
The data point for NGC 4486 (M87) is also highlighted. Here we use the galaxy R$_e$ from Forbes et al. (2017), based on Spitzer 3.6$\mu$m imaging, of around 7 kpc. However, we note that Kormendy et al.  (2009) found a much larger size of around 50 kpc. If correct, the latter would reduce the ratio for NGC 4486 to 1.7. 
Excluding NGC 4486, and the new HR17 data, we find a mean value for the ratio of 3.7 $\pm$ 0.4 for ETGs. This indicates that the galactocentric radius corresponding to 
half of the GC system is $\sim$4$\times$ larger than the radius containing half of the galaxy's light.  HR17 found a similar mean ratio, i.e $\sim$3.5. 
We also note a weak trend for the ratio to be higher in larger galaxies (as noted by HR17). Ultra Diffuse Galaxies and the Milky Way lie within the general scatter. 

\begin{figure}
      \includegraphics[angle=-90, width=0.5\textwidth]{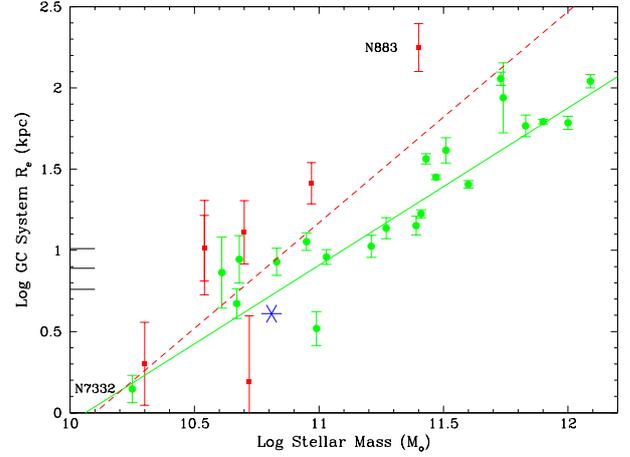}
	    \caption{\label{fig:corrGC2} 
Effective radius of the globular cluster system vs host galaxy stellar
mass. Red squares represent data from HR17 and
filled green circles data from the existing literature. 
The blue asterisk represents the Milky
Way's GC system, which follows the general trend. The 3 Ultra Diffuse Galaxies, all with log stellar masses around 8, are indicated by short horizontal lines.  
The green solid line shows a best fit relation between the GC system size and galaxy mass for the 
existing literature data for early-type galaxies (i.e. excluding the HR17 data, UDGs and the Milky Way). The slope of 0.97 $\pm$ 0.4 is 
fully consistent with a linear relation. 
The UDGs do not follow an extrapolation of the best fit to lower mass.
The red dashed line shows the fit of HR17 (slope = 1.30 $\pm$ 0.14) to their sample of 35 early and late-type galaxies. 
}

\end{figure} 

In Figure 2 we show GC system size vs host galaxy total stellar mass. NGC 4486, an outlier in Fig. 1, lies within the general scatter. A weighted best fit to the early-type galaxy data (excluding the new HR17 data and UDGs) gives:  log GC R$_e$ = 0.97 ($\pm$0.4)~log M$_{\ast}$ -- 9.76 ($\pm$4.4). 
This is fully consistent with a  linear trend between GC system size and the stellar mass of the host galaxy. This suggests that as early-type galaxies grow in stellar mass, their GC systems grow proportionally in size.  HR17 measured a slope of 1.30 $\pm$ 0.14 between GC R$_e$ and galaxy stellar mass for their sample of 35 early and late-type galaxies; thus the two slopes agree within the combined uncertainties.  The GC system of the Milky Way is consistent with the trend in Fig. 2. However the UDGs indicate that they follow a different scaling with stellar mass than a simple extrapolation to lower masses. We suspect that the relation flattens out for low galaxy masses with an inflection point around log M$_{\ast}$ = 
10.6 (i.e. at the same mass associated with the change in the slope of the galaxy size - stellar mass relation; Shen et al. 2003). We note that NGC 7332 (with log M$_{\ast}$ = 10.25) is a disturbed galaxy and the only one for which the quoted GC R$_e$ is less than the galaxy R$_e$; it warrants further study to confirm its GC system size.


\begin{figure}
      \includegraphics[angle=-90, width=0.5\textwidth]{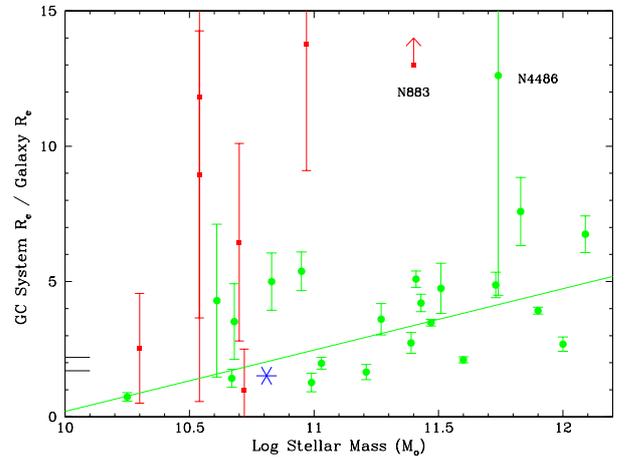}
	    \caption{\label{fig:corrGC2} The ratio of the size of a globular cluster system to its host galaxy vs galaxy stellar mass. 
The red filled squares are galaxies from the study of HR17; NGC 833 with a ratio of 46 is shown as a lower limit. 	    
The green filled circles are other data from the existing literature. Excluding the HR17 data and the UDGs, we find that the ratio of GC system to galaxy size increases for more massive galaxies. The green line shows a best fit of slope 2.27 $\pm$ 0.4. 
The blue asterisk represents the Milky
Way's GC system, which follows the general trend. The 3 Ultra Diffuse Galaxies, all with log stellar masses around 8, are indicated by short horizontal lines.  The UDGs do not follow an extrapolation of the best fit to lower mass.}

\end{figure} 

In Figure 3 we show the ratio of GC system to host galaxy size vs host galaxy stellar mass. Again excluding the new HR17 data from our analysis, we find that although the ratio has a mean value of around 4 it is larger for more massive galaxies. 
A weighted best fit relation has the form: ratio = 2.27 ($\pm$0.4)~log M$_{\ast}$ -- 22.5 ($\pm$4.4), where ratio = GC system R$_e$ / galaxy R$_e$. In low mass early-type galaxies, GC systems extend about 2-3$\times$ the effective radius of their host galaxies; for the highest mass galaxies in the universe the GC systems are even more extended at  5$\times$ the galaxy effective radius. Again, the GC system of the Milky Way obeys the early-type galaxy  relation but those of Ultra Diffuse Galaxies do not obey a simple extrapolation to log masses of $\sim$8. 

In the two-phase picture of ETG formation, more massive galaxies contain a larger fraction of accreted material from satellites (e.g. Oser et al. 2010). This accreted material serves to build-up the halo of the host galaxy, so that the more massive galaxies tend to have shallower, more extended  
surface brightness profiles (Pillepich et al. 2014). Figure 3 (and to some extent Figure 2) indicate that larger, more massive ETGs host more extended GC systems. This is consistent with the idea that 
GCs in the outer halos of ETGs are largely accreted from disrupted satellite galaxies (Georgiev et al. 2010; 
 Forbes et al. 2011; Blom et al. 2012). This may also indicate that massive galaxies accrete a larger fraction of low mass compared to high mass satellites which are disrupted at relatively large galactocentric radii (e.g. Oser et al. 2012).

\begin{figure}
      \includegraphics[angle=-90, width=0.5\textwidth]{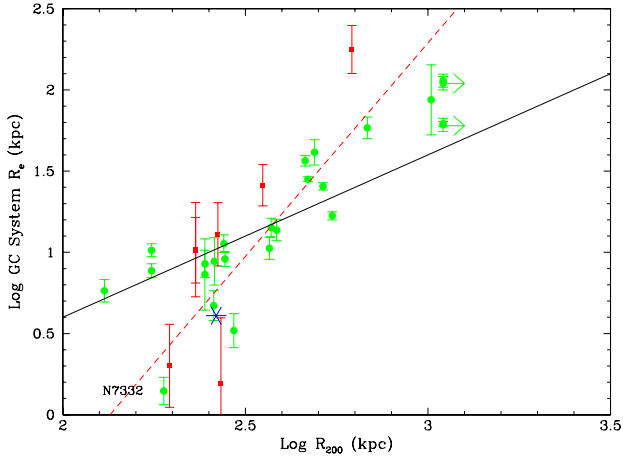}
	    \caption{\label{fig:corrGC2} Globular cluster system size vs virial radius (R$_{200}$) of the halo. 
The red filled squares are galaxies from the study of HR17. 
The green filled circles are other data from the existing literature. 
The blue asterisk represents the Milky
Way's GC system, which follows the general trend. The three galaxies with the smallest virial radii are the Ultra Diffuse Galaxies. 
The four most massive galaxies in our sample lack virial radii, and so are shown as lower limits.   The solid line is not a fit but shows the linear relation of Kravtsov (2013) between galaxy size and virial radius scaled up by a factor of 3.7 (i.e. the mean GC system to galaxy size ratio). The red dashed line is the fit of HR17 (slope = 2.63 $\pm$ 0.38) to their sample of 35 early and late-type galaxies.
}

\end{figure} 

HR17 used weak lensing results to connect their measured stellar masses to halo masses (M$_{200}$) and virial radii (R$_{200}$). Here we use values taken directly from their table 4.  For the UDGs we use the halo masses of 5$\times$10$^{11}$ M$_{\odot}$ for DF44 and DFX1 (van Dokkum et al. 2017), and 10$^{11}$ M$_{\odot}$ for DF17 (Peng et al. 2016), and assign approximate virial radii of 130 and 175 kpc respectively. The four most massive galaxies in our sample lack halo masses and virial radii. As they have stellar masses comparable to, or greater than NGC 4486, we show them as having virial radii and halo masses larger than NGC 4486 in the following figures. 

In Figure 4 we show the GC system size as function of the virial radius. The literature data show a general trend of increasing GC system size in larger halos, with the HR17 data having a large scatter. 
HR17 measured a slope of 2.63 $\pm$ 0.38, which is a reasonable representation for the GC systems in intermediate mass galaxies. However as can be seen in Fig. 4, their relation tends to overpredict the GC system size of the largest galaxies and underpredict those of UDGs. 
We also include in Fig. 4 the predicted linear relation 
from Kravtsov (2013), based on halo abundance matching, between the 2D projected 
galaxy effective radius and virial radius i.e. 
R$_e$ = 0.011 R$_{200}$ but scaled up by a factor of 3.7 to account for the typical GC system to galaxy size ratio. The resulting linear relation has a reasonable normalisation and slope compared to  the data, including the largest galaxies (which have lower limits on their size) and the smallest galaxies (the UDGs). 

Kravtsov (2013) argues that the near linear galaxy size-virial radius relation is consistent with the idea that galaxy sizes
are set by the angular momentum imparted to halos during formation  (e.g. Mo et al. 1998). 
Thus the size of a GC system (and the host galaxy), is to first order, set at early times in this model with 
subsequent evolution moving galaxies along the relation. 
A similar situation may exist for the linear relation between GC system mass and halo mass. For example,  
Boylan-Kolchin (2017) suggest that this relation is established at high redshift  with (metal-poor) GCs forming in direct proportion to the dark matter content of their host galaxy's halo and that mass growth over time maintains the linear relation.

Galaxy growth over time is a function of galaxy mass with more massive galaxies accreting a 
larger fraction of their mass compared to lower mass galaxies which are dominated by {\it in-situ} star formation (e.g. Oser et al. 2010; Pillepich et al. 2014). 
These two modes of stellar mass growth are also relevant for GC systems, with GCs expected to form 
in-situ and to be accreted along satellite galaxies. The relative importance of in-situ to accreted GCs would be expected to vary as a function of host galaxy mass, as would tidal stripping and the destruction of GCs. Future simulations, that include these processes, will help our understanding of  how such linear relations can be maintained over time.

\begin{figure}
      \includegraphics[angle=-90, width=0.5\textwidth]{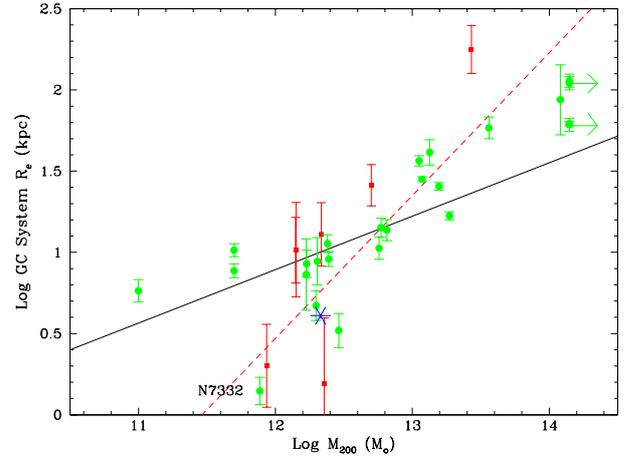}
	    \caption{\label{fig:corrGC2} Globular cluster system size vs the halo mass (M$_{200}$). 
The red filled squares are galaxies from the study of HR17. 
The green filled circles are other data from the existing literature. 
The blue asterisk represents the Milky
Way's GC system, which follows the general trend. The three galaxies with the smallest halo masses are the Ultra Diffuse Galaxies. The red dashed line is the fit of HR17 (slope = 0.88  $\pm$ 0.10) to their sample of 35 early and late-type galaxies.
The solid line is not a fit but shows a relation of slope 1/3 based on the predictions of Kravtsov (2013). 
}

\end{figure}

In Figure 5 we show the GC system size as a function of halo mass. HR17 found a slope of 0.88 $\pm$ 0.10 for their sample, which again provides a reasonable representation for our 
intermediate halo mass galaxies.  However, our inclusion of higher mass early-type galaxies (with lower limits on their halo mass) and three lower mass UDGs suggests that the actual relation is much shallower. In Fig. 4 we found that the slope of the GC system size vs R$_{200}$ is around unity. The virial, or halo, mass M$_{200}$ $\propto$  R$_{200}^3$. So based on Fig. 4, and the predictions of Kravtsov (2013), we would expect GC system size to scale with M$_{200}^{1/3}$. A relation of this slope is included in Fig. 5 and it indeed provides a 
good representation of the data over the full mass range. This suggests that GC system size does not scale with M$_{200}^{0.88}$ but closer to M$_{200}^{0.33}$ and that the most massive galaxies in the universe and UDGs follow this scaling relation. 

\section{Conclusions}

Here we have examined the size of GC systems of 22 early-type galaxies using a measure of their half number effective radii from fits to GC surface density profiles. We exclude late-type galaxies and  the most recent new data for 7 GC systems from Hudson \& Robison (2017) as this new data shows strong deviations from the existing data. Our analysis extends to lower and higher masses compared to Hudson \& Robison. We find 
a linear relation between the GC system size and its host galaxy stellar mass but there are indications from Ultra Diffuse Galaxies that the relation may flatten for log stellar masses less than 10.6 M$_{\odot}$. 
We measure the ratio of the GC system size to that of its host galaxy, finding a mean value of 3.7 for our sample. However, this ratio increases from around 3$\times$ for M$^{\ast}$ galaxies to around 5$\times$ for the most massive galaxies in the universe. 

Our main result is that GC system size has an approximately  linear relation with virial radius and halo mass to the 1/3 power. This is consistent with the galaxy scalings predicted by Kravtsov (2013) and suggests that the relation is set during the initial phases of galaxy formation. 
Thus complementary to the known linear scaling of GC system mass with halo mass, we also find  a near linear scaling of GC system size with virial radius. These indicate a strong connection between GC systems and host galaxy dark matter properties. The GC system of the Milky Way appears to follow the same scalings as the GC systems of early-type galaxies. 
The GC system sizes of UDGs also scale with virial radius and halo mass in the same sense as more massive, larger  galaxy halos. 
Our larger host galaxy mass range has revealed different scalings of GC system size with virial radius and halo mass  to those claimed by Hudson \& Robison (2017) , but these scalings should still be verified with a more homogeneous sample.
Future work should also attempt to bridge the gap in mass between UDGs and massive early-type galaxies, and study  the blue and red GC subpopulations independently.

\section{Acknowledgements}

DAF thanks the ARC for financial support via DP130100388 and SLUGGS survey team members. We thank A. Romanowsky for several useful suggestions. We thank the University of Surrey for providing such a welcoming environment where this work was carried out and the Santander Fellowship programme for their support. We thank the referee for several useful suggestions.

\section{References}

Alabi A.~B., et al., 2016, MNRAS, 460, 3838\\
Alabi A.~B., et al., 2017, MNRAS, 468, 3949\\
Boylan-Kolchin M., 2017, arXiv, arXiv:1705.01548\\
Blakeslee J.~P., 1997, ApJ, 481, L59\\
Blom C., Spitler L.~R., Forbes D.~A., 2012, MNRAS, 420, 37\\
Brodie J.~P., et al., 2014, ApJ, 796, 52 \\
Cappellari M., et al., 2011, MNRAS, 413, 813 \\
Forbes D.~A., Spitler L.~R., Strader J., Romanowsky A.~J., Brodie J.~P., Foster C., 2011, MNRAS, 413, 2943 \\
Forbes D.~A., Sinpetru L., Savorgnan G., Romanowsky A.~J., Usher C., Brodie J., 2017, MNRAS, 464, 4611 \\
Georgiev I.~Y., Puzia T.~H., Hilker M., Goudfrooij P., 2010, MNRAS, 409, 447\\
Hargis J.~R., Rhode K.~L., 2012, AJ, 144, 164\\
Hargis J.~R., Rhode K.~L., 2014, ApJ, 796, 62 \\
Harris W.~E., Harris G.~L., Hudson M.~J., 2015, ApJ, 806, 36 \\
Harris W.~E., Blakeslee J.~P., Harris G.~L.~H., 2017a, ApJ, 836, 67 \\
Harris W.~E., Ciccone S.~M., Eadie G.~M., Gnedin O.~Y., Geisler D., Rothberg B., Bailin J., 2017b, ApJ, 835, 101 \\
Hudelot, P. et al. 2012, VizieR Online Data Catalog, 2317, 0\\
Hudson, M., Robison, B., 2017, arXiv:1707.02609v1, MNRAS submitted\\
Kartha S.~S., Forbes D.~A., Spitler L.~R., Romanowsky A.~J., Arnold J.~A., 
Brodie J.~P., 2014, MNRAS, 437, 273 \\
Kartha S.~S., et al., 2016, MNRAS, 458, 105\\
Kormendy J., Fisher D.~B., Cornell M.~E., Bender R., 2009, ApJS, 182, 216 \\
Kravtsov A.~V., 2013, ApJ, 764, L31\\
McMillan P.~J., 2011, MNRAS, 414, 2446 \\
Mo H.~J., Mao S., White S.~D.~M., 1998, MNRAS, 295, 319\\
Oser L., Ostriker J.~P., Naab T., Johansson P.~H., Burkert A., 2010, ApJ, 725, 2312 \\
Oser L., Naab T., Ostriker J.~P., Johansson P.~H., 2012, ApJ, 744, 63 \\
Peng E.~W., et al., 2011, ApJ, 730, 23 \\
Peng E.~W., Lim S., 2016, ApJ, 822, L31 \\
Pillepich A., et al., 2014, MNRAS, 444, 237\\
Pota V., et al., 2013, MNRAS, 428, 389 \\
Rhode K.~L., Zepf S.~E., Kundu A., Larner A.~N., 2007, AJ, 134, 1403 \\
Rhode K.~L., Windschitl J.~L., Young M.~D., 2010, AJ, 140, 430 \\
Shen S., Mo H.~J., White S.~D.~M., Blanton M.~R., Kauffmann G., Voges W., Brinkmann J., Csabai I., 2003, MNRAS, 343, 978 \\
Spitler L.~R., Romanowsky A.~J., Diemand J., Strader J., Forbes D.~A., Moore B., Brodie J.~P., 2012, MNRAS, 423, 2177 \\
Spitler L.~R., Forbes D.~A., 2009, MNRAS, 392, L1 \\
Usher C., et al., 2012, MNRAS, 426, 1475 \\
Usher C., Forbes D.~A., Spitler L.~R., Brodie J.~P., Romanowsky A.~J., 
Strader J., Woodley K.~A., 2013, MNRAS, 436, 1172 \\
van den Bergh S., 2000, PASP, 112, 932 \\
van Dokkum P., Conroy C., Villaume A., Brodie J., Romanowsky A.~J., 2017, ApJ, 841, 68 \\
Veale M., et al., 2017, MNRAS, 464, 356 \\
Vika M., Driver S.~P., Cameron E., Kelvin L., Robotham A., 2012, MNRAS, 419, 2264 \\
Young M.~D., Dowell J.~L., Rhode K.~L., 2012, AJ, 144, 103 \\
\end{document}